\documentclass[aps,longbibliography,pre, reprint,groupaddress,bibnotes]{revtex4-1}
\usepackage{graphicx}
\usepackage{epstopdf}

\usepackage{dcolumn}
\usepackage{bm, amssymb, color}
\usepackage{hyperref}
\usepackage{natbib}
\usepackage{booktabs,tabulary}
\usepackage{multirow}
\usepackage{amsmath,soul}
\usepackage[export]{adjustbox}
\usepackage{float}
\usepackage[caption=false]{subfig}
\usepackage{array}
\usepackage{cases}
\usepackage{makecell}
\usepackage{physics}
\usepackage{yhmath}
\usepackage{enumitem}

\newcommand{\spD}[1]{\fn{\tilde{\chi}_{_V}}{#1}}

\newcommand{\tens}[1]{\mathinner{\boldsymbol{#1}}}

\newcommand{\uvect}[1]{\hat{\vect{#1}}}

\setstcolor{blue}

\setul{0}{0.3ex}

\newcommand{\R}{\mathbb{R}}
\renewcommand{\d}[1]{\mathinner{d#1}}
\newcommand{\fn}[2]{\mathinner{#1\mathopen{\left(#2\right)}}}
\newcommand{\vect}[1]{{\bf #1}}

\usepackage[colorinlistoftodos,shadow,textwidth=18mm]{todonotes}
\newcommand{\vv}[1]{\fn{\sigma_V ^2}{#1}}

\usepackage{stackengine}
\setstackgap{S}{1pt}

\begin{document}

\title{Characterizing the Hyperuniformity
of Ordered and Disordered Two-Phase Media}

\author{Jaeuk Kim}
\affiliation{Department of Physics, Princeton University, Princeton, New Jersey 08544, USA}
\author{Salvatore Torquato}%
\email{torquato@princeton.edu}
\homepage{http://torquato.princeton.edu}
\affiliation{Department of Physics, Princeton University, Princeton, New Jersey 08544, USA}
\affiliation{Department of Chemistry, Princeton University, Princeton, New Jersey 08544, USA}
\affiliation{Princeton Institute for the Science and Technology of Materials, Princeton University, Princeton, New Jersey 08544, USA}
\affiliation{Program in Applied and Computational Mathematics, Princeton University, Princeton, New Jersey 08544, USA}

\date{\today}

\begin{abstract}

The hyperuniformity concept provides a unified means to classify all perfect crystals, perfect quasicrystals, and exotic amorphous 
states of matter according to their capacity to suppress large-scale density fluctuations. 
While the classification of hyperuniform point configurations has received considerable attention, much less is known about the classification of  hyperuniform heterogeneous two-phase media, which include composites, porous media, foams, cellular solids, colloidal suspensions and polymer blends. 
The purpose of this article is to begin such a program for certain two-dimensional models of hyperuniform two-phase media by ascertaining their local volume-fraction variances 
$\sigma^2_{_V}(R)$ and the associated hyperuniformity order metrics $\overline{B}_V$.
This is a highly challenging task because the geometries and 
topologies of the phases are generally much richer and more complex than point-configuration arrangements and one must ascertain a broadly applicable length scale to make key quantities dimensionless. Therefore, we purposely restrict ourselves to a certain class 
of two-dimensional periodic cellular networks as well as periodic and disordered/irregular packings, 
some of which maximize their effective transport and elastic properties. 
Among the cellular networks considered, the honeycomb networks have the minimal 
value of the hyperuniformity order metrics $\overline{B}_V$ across all volume fractions. 
On the other hand, among all packings considered, the triangular-lattice packings have the smallest values of $\overline{B}_V$ for the possible range of volume fractions.
Among all structures studied here, the triangular-lattice packing has the minimal order metric for almost all volume fractions.
Our study provides a theoretical foundation for the establishment of hyperuniformity order metrics for general two-phase media and a basis to discover new hyperuniform two-phase systems by inverse design procedures.

\end{abstract}

\date{\today}
\maketitle
\section{Introduction}

The hyperuniformity concept generalizes the traditional notion of long-range order
in many-particle systems to include
all perfect crystals, perfect quasicrystals, and exotic amorphous states
of matter \cite{Torquato2003_hyper,Torquato2018_review}. A hyperuniform point configuration in $d$-dimensional Euclidean space $\mathbb{R}^d$  is characterized by an anomalous suppression
of large-scale density fluctuations relative to those in typical disordered systems, such as liquids and structural glasses.
The hyperuniformity notion was generalized to the case of heterogeneous (multiphase) materials \cite{Zachary2009a,  Torquato2016_gen, ma_precise_2018}, i.e., materials consisting of two or more phases \cite{Torquato_RHM, Sahimi_HM1}, such as composites, porous media, foams, cellular solids, colloidal suspensions and polymer blends.
Subsequently, the concept was extended to quantify hyperuniformity in a variety of different systems, including random scalar fields, divergence-free random vector fields, and statistically anisotropic many-particle systems \cite{Torquato2016_gen}. Hyperuniformity has been attracting great attention across many fields, including physics  \cite{Torquato2015_stealthy, zhang_perfect_2016, Hexner2017_2, Ricouvier2017, Oguz2017, Ma2017, Lopez2018, yu_disordered_2018, wang_hyperuniformity_2018, Torquato2018_review, lei_hydrodynamics_2019, gorsky_engineered_2019, Gabrielli2004, Jaeuk2018, klatt_cloaking_2020, Torquato2018_review}, materials science  \cite{ma_3d_2016, xu_microstructure_2017, torquato_multifunctional_2018, Chen2017, kim_new_2019}, mathematics  \cite{Sodin2006, Peres2014, Ghosh2017_holeConjecture, brauchart_hyperuniform_2019, torquato_hidden_2019} and biology \cite{Jiao2014_chickenEyes, Mayer2015, Torquato2018_review, zheng_hyperuniformity_2020}.

In the case of point configurations, one can rank order crystals, quasicrystals, and special disordered systems within a {\it hyperuniformity class} according to the degree to which they suppress density fluctuations, as measured by the hyperuniformity parameter $\overline{B}_N$ \cite{Torquato2003_hyper, Zachary2009a}.
Much less is known about the analogous rank ordering of hyperuniform two-phase media via the appropriate hyperuniformity parameter $\overline{B}_V$, as defined below.
However, it is much more challenging to do so for two-phase media  for two reasons.
First, the geometries and topologies of the phases are generally much richer and more complex than point-configuration arrangements. 
Second, one must determine length scales that are broadly applicable for the multitude of possible two-phase media microstructures to make $\overline{B}_V$ dimensionless. 
The purpose of this article is to begin such a program for certain two-dimensional periodic and disordered models of two-phase media.

For two-phase heterogeneous media in $d$-dimensional Euclidean space $\R^d$, which include cellular solids, composites, and porous media, hyperuniformity is defined by the following infinite-wavelength  condition on the {\it spectral
density} $\spD{\vect{k}}$\cite{Zachary2009a, Torquato2018_review}, i.e., 
\begin{equation}\label{eq:HU_condition}
\lim_{\abs{\vect{k}}\to 0 }\spD{\vect{k}} = 0,
\end{equation}
where $\vect{k}$ is the wavevector.
The spectral density $\spD{\vect{k}}$ is the Fourier transform of the autocovariance function $\fn{\chi_{_V}}{\vect{r}}\equiv \fn{S_2^{(i)}}{\vect{r}}-{\phi_i}^2$, where $\phi_i$ is the volume fraction of phase $i$, and $\fn{S_2^{(i)}}{\vect{r}}$ gives the probability of finding two points separated by $\vect{r}$ in phase $i$ at the same time \cite{Sahimi_HM1, Torquato_RHM}.
It can be easily obtained  for general microstructures either theoretically, computationally, or via scattering experiments \cite{Debye1949}.
Hyperuniformity can equivalently be defined in terms of the {\it local volume-fraction variance} $\vv{R}$ associated with a spherical window of radius $R$. 
Specifically, a hyperuniform two-phase system is one in which $\vv{R}$ decays faster than $R^{-d}$ in the large-$R$ regime \cite{Zachary2009a, Torquato2018_review}, i.e.,  
\begin{equation}\label{eq:HU-condition2}
\lim_{R\to\infty}R^d \vv{R} = 0.
\end{equation}
The local variance $\vv{R}$ is directly determined by $\spD{\vect{k}}$ via the following integral \cite{Torquato2018_review, Zachary2009a}: 
\begin{align}
\vv{R} & = \frac{1}{\fn{v_1}{R}(2\pi)^d} \int_{\R^d} \spD{\vect{k}} \fn{\tilde{\alpha}_2}{\abs{\vect{k}};R} \dd{\vect{k}}
, \label{eq:vv-Fourier}
\end{align}
where $\fn{v_1}{R}=\pi^{d/2}[\fn{\Gamma}{d/2+1}]^{-1} R^d$ is the volume of a $d$-dimensional sphere of radius $R$, $\fn{\Gamma}{x}$ is the gamma function, and 
\begin{equation} \label{eq:alpha-tilde}
\fn{\tilde{\alpha}_2}{\abs{\vect{k}};R} \equiv 2^d \pi^{d/2} \fn{\Gamma}{d/2+1}\frac{\qty[\fn{J_{d/2}}{\abs{\vect{k}}R}]^2}{\abs{\vect{k}}^d},
\end{equation} 
is the Fourier transform of the scaled intersection volume of two spheres of radius $R$ that are separated by $r$.

As in the case of hyperuniform point configurations  \cite{Torquato2003_hyper, Zachary2009a, Zachary2011_2, Torquato2018_review},  there are three different scaling regimes (classes) that describe the associated large-$R$ behaviors of the volume-fraction variance when the spectral density goes to zero as a power-law scaling  ${\tilde \chi}_{_V}({\bf Q})\sim |{\bf Q}|^\alpha$ as $Q$ tends to zero:
\begin{align}  
\sigma^2_{_V}(R) \sim 
\begin{cases}
R^{-(d+1)}, \quad\quad\quad \alpha >1 \qquad &\text{(Class I)}\\
R^{-(d+1)} \ln R, \quad \alpha = 1 \qquad &\text{(Class II)},\\
R^{-(d+\alpha)}, \quad 0 < \alpha < 1\qquad  &\text{(Class III)}
\end{cases}
\label{eq:classes}
\end{align}
where the exponent $\alpha$ is a positive constant.
Classes I and III are the strongest and weakest forms of hyperuniformity, respectively.
One aim of this paper is to compute the implied coefficient hyperuniformity order metric $\overline{B}_V$ (defined in Sec. \ref{sec:asy}) multiplying $R^{-(d+1)}$ for certain class I structures, which is a measure of the degree to which large-scale volume-fraction fluctuations are suppressed within that class. 

An overarching goal of this paper is to characterize the hyperuniformity of models of two-phase media that belong to class I.
Due to the infinite variety of possible two-phase microstructures (geometries and topologies), we purposely restrict ourselves to a certain class of periodic cellular networks as well as periodic and disordered/irregular packings.
Even this restrictive set of models of two-phase media presents challenges, since one must ascertain relevant length scales that are broadly applicable to make key quantities dimensionless, as discussed in Sec. \ref{sec:length-scale}.
In particular, we evaluate the volume-fraction variance as a function of the window radius $R$ for all models.
We also compute the aforementioned hyperuniformity order metric $\overline{B}_V$ for each model to rank order them.

In Sec. \ref{sec:models}, we present relevant theoretical background
to characterize hyperuniform two-phase media and describe the computational methods employed in this study. 
In Sec. \ref{sec:form-factor}, we provide exact closed-form formulas of the form factors of general polyhedra in $\R^2$ and $\R^3$, which are important to characterize periodic networks.
We then describe the two-dimensional models of two-phase media of class I hyperuniformity considered in this investigation: periodic cellular networks (Sec. \ref{sec:models}), periodic disk packings, and disordered/irregular disk packings (Sec. \ref{sec:packings}). 
In Sec. \ref{sec:length-scale}, we provide the rationale for choosing the inverse of the specific surface as the characteristic length scale $D$ in the hyperuniformity order metric $\overline{B}_V$. 
In Sec. \ref{sec:results}, we investigate the microstructure-dependence of the volume-fraction variance and rank order all of our class I models according to $\overline{B}_V$.
Finally, we present concluding remarks and outlooks for future research in Sec. \ref{sec:conclusion}.

\section{Background and Methods}\label{sec:prel}

\subsection{Asymptotic Analysis of Hyperuniform Two-Phase Media}
\label{sec:asy}

For a statistically homogeneous and isotropic medium in $\R^d$, the local volume-fraction variance $\vv{R}$ can be written as the following large-$R$ asymptotic expansion \cite{Zachary2009a, Torquato2018_review}:
\begin{equation}\label{eq:large-R_exp}
\vv{R} = \fn{A_V}{R} \qty(\frac{D}{R})^d + \fn{B_V}{R} \qty(\frac{D}{R})^{d+1} + o\qty(\frac{D}{R})^{d+1},
\end{equation}
where $\fn{A_V}{R}$ and $\fn{B_V}{R}$ are dimensionless asymptotic coefficients of powers $R^{-d}$ and $R^{-(d+1)}$, respectively, and they are defined by 
\begin{align}
\fn{A_V}{R} =& \frac{1}{\fn{v_1}{D}} \int_{\abs{\vect{r}}\leq 2R} \fn{\chi_{_V}}{\vect{r}}\dd{\vect{r}} \\
\fn{B_V}{R} =& -\frac{\fn{c}{d}}{2D\fn{v_1}{D}} \int_{\abs{\vect{r}}\leq 2R} \fn{\chi_{_V}}{\vect{r}}\abs{\vect{r}}\dd{\vect{r}},
\end{align}
where $\fn{c}{d}\equiv 2\fn{\Gamma}{d/2+1}/\qty[\pi^{1/2} \fn{\Gamma}{(d+1)/2}]$, and $D$ is a characteristic length scale of the medium.
In the large-$R$ limit, the coefficient $\fn{A_V}{R}$ is proportional to the spectral density at the origin, i.e.,
\begin{equation}
\overline{A}_V \equiv \lim_{R\to\infty}\fn{A_V}{R} \propto \lim_{\abs{\vect{k}}\to 0} \spD{\vect{k}},
\end{equation}
and thus for any hyperuniform medium, $\overline{A}_V=0$, and hence the expansion \eqref{eq:large-R_exp} reduces to
\begin{equation}
\vv{R} = \fn{B_V}{R} \qty(\frac{D}{R})^{d+1} + o\qty(\frac{D}{R})^{d+1} .
\end{equation} 
It is noteworthy that, unlike $\vv{R}$, the coefficient $\fn{B_V}{R}$ depends on the choice of the length scale $D$. 

In the case of class I hyperuniform systems, $\vv{R}$ decays like $R^{-(d+1)}$ for large $R$, as specified by 
\begin{equation} \label{eq:vv-asy}
\vv{R}\sim \overline{B}_V \qty(\frac{D}{R})^{d+1},\qquad R\to\infty.
\end{equation}
As $R$ increases, the coefficient $\fn{B_V}{R}$ converges to the hyperuniformity order metric $\overline{B}_V$ for typical disordered systems.
For some infinite media, such as periodic and aperiodic structures,the associated coefficient $\fn{B_V}{R}$ oscillates around some running average value. 
In such cases, it is advantageous to estimate $\overline{B}_V$ by using the cumulative moving average, as defined by \cite{Torquato2018_review}
\begin{equation}\label{eq:Bv-asy}
\overline{B}_V \equiv \lim_{L\to\infty} \frac{1}{L} \int_0^L \fn{B_V}{R} \dd{R}.
\end{equation}


\subsection{Spectral Density and the Local Volume-Fraction Variance} \label{sec:methods}

%

Here we present explicit formulas for the spectral density $\spD{\vect{k}}$ of general packings in $\R^d$, ordered or not \cite{zachary_hyperuniformity_2011, torquato_disordered_2016, Torquato2018_review}.
We also describe the formula for the local volume-fraction variance $\vv{R}$ for class I hyperuniform two-phase media and the associated hyperuniformity order metric $\overline{B}_V$.
Importantly, these formulas also can be applied to characterize periodic cellular networks, as we will discuss later. 

In the case of packings of identical particles $\tens{P}$ of arbitrary shape, it is known that 
\begin{equation}\label{eq:chik-mono}
\spD{\vect{k}} = \rho \abs{\fn{\tilde{m}}{\vect{k};\tens{P}}}^2 \fn{S}{\vect{k}},
\end{equation}
where $\rho$ is the number density of particle centers, $\fn{\tilde{m}}{\vect{k};\tens{P}}$ is the Fourier transform of (also called {\it form factor}) of the particle indicator function $\fn{m}{\vect{x};\tens{P}}$ defined by
\begin{equation}
\fn{m}{\vect{x};\tens{P}} 
=
\begin{cases}
1, &\vect{x} \text{ is inside } \tens{P}\\
0, &\text{otherwise}
\end{cases},
\end{equation}
where $\vect{x}$ is the position vector with respect to the centroid of $\tens{P}$, and $\fn{S}{\vect{k}}$ is the structure factor of particle centers; see Refs. \cite{zachary_hyperuniformity_2011, torquato_disordered_2016, Torquato2018_review} for the definition of $\fn{S}{\vect{k}}$ and its computation. 
One can immediately obtain from \eqref{eq:chik-mono} the specific formulas for a $d$-dimensional (Bravais) lattice packing in which a single particle $\tens{P}$ is placed in a fundamental cell $\mathcal{F}$ of the Bravais lattice $\mathcal{L}$ as follows:
\begin{align}\label{eq:chik-network-1pt}
\spD{\vect{k}} &= {\abs{V_\mathcal{F}}}^{-1} \abs{\fn{\tilde{m}}{\vect{k};\tens{P}} }^2 \fn{S_\mathcal{L}}{\vect{k}},  
\end{align}
where $\abs{V_\mathcal{F}}$ is the volume of the fundamental cell $\mathcal{F}$ of $\mathcal{L}$, $\fn{S_\mathcal{L}}{\vect{k}}$ is the structure factor of $\mathcal{L}$ given by \cite{Torquato2018_review}
\begin{equation}
\fn{S_\mathcal{L}}{\vect{k}} = \frac{(2\pi)^d}{\abs{V_\mathcal{F}}} \sum_{\vect{q}\in \mathcal{L}^* \setminus\{\vect{0}\}} \fn{\delta}{\vect{k}-\vect{q}},
\end{equation}
$\mathcal{L}^*$ denotes the reciprocal lattice of $\mathcal{L}$, and $\fn{\delta}{\vect{x}}$ is the Dirac delta function.
For a periodic packing in which a fundamental cell contains $M$ distinct particles ($\tens{P}_1,\cdots, \tens{P}_M$) whose centroids are at $\vect{r}_1,\cdots,\vect{r}_M$, formula \eqref{eq:chik-network-1pt} can be easily extended as 
\begin{align}\label{eq:chik-network-M}
\spD{\vect{k}} &= {\abs{V_\mathcal{F}}}^{-1} \abs{ \fn{\tilde{m}}{\vect{k};\qty{\tens{P}_j}} }^2 \fn{S_\mathcal{L}}{\vect{k}},  
\end{align}
where 
\begin{equation}\label{eq:form-factor}
\fn{\tilde{m}}{\vect{k};\qty{\tens{P}_j}} \equiv \sum_{j=1}^M \fn{\tilde{m}}{\vect{k};\tens{P}_j} e^{-i\vect{k}\cdot\vect{r}_j}.
\end{equation}
Equation \eqref{eq:chik-network-M} is a special case of the multicomponent packing formula given in Ref. \cite{torquato_disordered_2016}.
Thus, given the form factors and structure factors for a particulate two-phase structure, one can immediately compute the corresponding spectral density via either \eqref{eq:chik-mono}, \eqref{eq:chik-network-1pt}, or \eqref{eq:chik-network-M}.
It is crucial to note that these formulas also can be applied to any periodic cellular network by treating it as a periodic packing of polygons (polyhedra for $d=3$) defined by the void regions (shown in white in Fig. \ref{fig:sch}).
In such cases, the set $\qty{\tens{P}_j}$  represents the regions of void phase (shown in white regions in Fig. \ref{fig:sch}).

%
%

Given the spectral density of a general packing, one can compute the local volume-fraction variance by computing Eq. \eqref{eq:vv-Fourier} either numerically or analytically.
The associated hyperuniformity order metric $\overline{B}_V$ is obtained from the running average associated with \eqref{eq:Bv-asy}.
In the case of periodic packings or periodic networks, it immediately follows from Eqs. \eqref{eq:vv-Fourier} and \eqref{eq:chik-network-M} that the associated local volume-fraction variance $\vv{R}$ and the surface-area coefficient $\fn{B_V}{R}$ are written as 
\begin{align}
\vv{R} 
	=& \frac{2^d \fn{\Gamma}{d/2+1}^2 }{R^d} \frac{1}{\abs{V_\mathcal{F}}^2} 
	\nonumber \\
	&\times\sum_{\vect{k}\in \mathcal{L}^* \setminus\{\vect{0}\}} \abs{\fn{\tilde{m}}{\vect{k};\qty{\tens{P}_j}}}^2 \frac{[\fn{J_{d/2}}{kR}]^2}{k^d} \label{eq:vv-perio-network}\\
	\sim& \fn{B_V}{R}\qty(\frac{D}{R})^{d+1},\qquad(R\to\infty).
	\label{eq:Bv-perio-network}
\end{align}
Thus, we see that periodic packings fall in class I. 
The hyperuniformity order metric $\overline{B}_V$ is obtained by substituting \eqref{eq:Bv-perio-network} into \eqref{eq:Bv-asy}: 
\begin{align}
\overline{B}_V &=\frac{2^d  \fn{\Gamma}{d/2+1}^2 }{\pi D^{d+1}} \frac{1}{\abs{V_\mathcal{F}}^2} \sum_{\vect{k}\in \mathcal{L}^* \setminus\{\vect{0}\} } \frac{\abs{\fn{\tilde{m}}{\vect{k};\qty{\tens{P}_j}}}^2}{q^{d+1}} \label{eq:Bv-asy-period},
\end{align}
where we have used the identity $
\lim_{x\to\infty}x^{-1}\int_0^{x} \d{x'} x' [\fn{J_{d/2}}{x'}]^2 = 1/\pi$.


%
\subsection{Computation of $\vv{R}$ and $\overline{B}_V$} 

Here we describe two methods that we employ to estimate the local volume-fraction variance $\vv{R}$ and the associated asymptotic value $\overline{B}_V$: numerical integration of Eq. \eqref{eq:vv-Fourier} and the Monte Carlo (MC) method.
For periodic media, we mainly use the former method because of its accuracy and efficiency for such  structures [cf. Eq. \eqref{eq:Bv-perio-network}]. 
The key step of this method is to compute the spectral density $\spD{\vect{k}}$ of a periodic structure via either \eqref{eq:chik-network-1pt} and \eqref{eq:chik-network-M}.
For periodic packings of identical circular disks,we use the exact formula for the form factor of a $d$-dimensional sphere of radius $a$ given by \cite{Torquato_RHM}
\begin{equation}
\fn{\tilde{m}}{k;a} = \qty(\frac{2\pi a}{k})^{d/2}{\fn{J_{d/2}}{ka}}.
\end{equation}
In the case of periodic networks, we employ the formulas for general polyhedra in two and three dimensions given in Sec. \ref{sec:form-factor}.
Provided that $\spD{\vect{k}}$ given in Eq. \eqref{eq:chik-network-M} can be computed, it is in practice sufficient to perform the summations in Eqs. \eqref{eq:Bv-perio-network} and \eqref{eq:Bv-asy-period} up to $\abs{\vect{k}}\abs{V_\mathcal{F}}^{1/d} < 1000$ for $d=2,3$.

Because the numerical calculations of \eqref{eq:vv-Fourier} can be computationally expensive, we employ the MC method to estimate $\vv{R}$ for disordered disk packings.
Specifically, $\vv{R}$ is estimated by uniformly sampling the local volume fraction with a $d$-dimensional spherical observation window of radius $R$ a single packing or an ensemble of packings.
Since this method involves computing the volume of domains in one phase intersected by a window, it is highly nontrivial and computationally expensive for general packings.
In the case of disk packings (sphere packings for $d=3$), however, such a calculation can be efficiently carried out by using an exact closed-form formula for the intersection volume of two spheres of radii $R_1$ and $R_2$ whose centers are separated by $r$, given in Ref. \cite{Torquato_RHM}.

\begin{figure}
\includegraphics[width=0.45\textwidth]{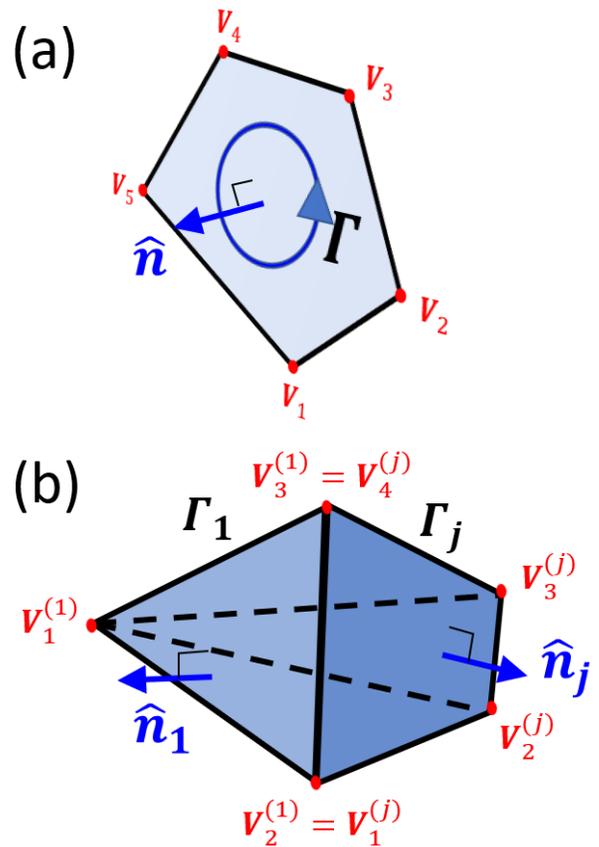}
\caption{Illustrations of parameters used to compute the form factors of polygonal figures in $\R^3$. 
(a) A pentagon has a face $\Gamma$ surrounded by five vertices $\vect{V}_1,\cdots,\vect{V}_5$.
The vertex indices increase counter-clockwise when the normal vector $\uvect{n}$ is towards the reader.
(b) A polyhedron with five faces, for each of which the ordering of vertices fulfills the right-hand-rule with the normal vector $\uvect{n}_j$.
Note that all normal vectors point towards the outside of the polyhedron.
\label{fig:polyhedra}}
\end{figure}

\section{Form Factors of Polygons and Polyhedra}
\label{sec:form-factor}
 
In order to compute the spectral density of a periodic cellular network using relation \eqref{eq:chik-network-M}, one needs to compute the form factors of the relevant polyhedra.
Here, we provide the exact closed-form formulas for general polyhedra in three dimensions and two dimensions (polygons) that were derived in Ref. \cite{wuttke_form_2017}.

We first consider a planar polygon $\tens{\Gamma}$ that is placed in an arbitrary orientation in $\R^3$ [see Fig. \ref{fig:polyhedra}(a)] and consists of $J$ vertices $\vect{V}_1, \cdots, \vect{V}_J$ in a cyclic order, implying that the adjacent vertices $\vect{V}_{i-1}$ and $\vect{V}_i$ are connected by a segment, and $\vect{V}_{J+1} = \vect{V}_1$.
It is convenient to consider a planar polygon in three dimensions since such planar polygons will be employed to define a polyhedron in $\R^3$ later.
For two adjacent vertices $\vect{V}_{i-1}$ and $\vect{V}_{i}$, we define 
\begin{align}
\vect{R}_i \equiv& \frac{1}{2}\qty(\vect{V}_i +\vect{V}_{i-1}), \\
\vect{E}_i \equiv& \frac{1}{2}\qty(\vect{V}_i -\vect{V}_{i-1}),
\end{align}
where $\vect{R}_i$ is the center of the two vertices, and $\vect{E}_i$ stands for the segment from $\vect{V}_{i-1}$ to $\vect{R}_i$.
The form factor of $\tens{\Gamma}$ at a wavevector $\vect{k}$ is
\begin{equation}\label{eq:form-factor_J-gon}
\fn{\tilde{m}}{\vect{k};\tens{\Gamma}} = \frac{2}{-i {k_{\parallel}}^2} \vect{k}_\times \cdot \sum_{j=1}^J \vect{E}_j \fn{\mathrm{sinc}}{\vect{k}\cdot \vect{E}_j}  e^{-i\vect{k}\cdot\vect{R}_j},
\end{equation}
where $\uvect{n}$ is the unit normal vector of the face $\tens{\Gamma}$, $\vect{k}_\parallel \equiv \vect{k}-(\vect{k}\cdot \uvect{n})\uvect{n}$, $\vect{k}_\times = \uvect{n}\times \vect{k}$, and 
\begin{equation}
\fn{\mathrm{sinc}}{x}\equiv
\begin{cases}
1, & x=0, \\
\frac{\sin{x}}{x}, &\mathrm{otherwsie}
\end{cases}.
\end{equation}
Importantly, the ordering of vertices should fulfill the {\it right-hand-rule} with respect to the normal vector $\uvect{n}$ [see Fig. \ref{fig:polyhedra}(a)], implying that the vertex index increases counterclockwise when $\uvect{n}$ is towards the reader. 
In two-dimensional applications, one should take $\vect{k} = \vect{k}_\parallel$.

We now consider a polyhedron $\tens{\mathcal{P}}$ with $K$ faces ($\tens{\Gamma}_1, \cdots, \tens{\Gamma}_K$) in which a face $\tens{\Gamma}_j$ is a polygon with $J_j$ vertices.
For each face $\tens{\Gamma}_j$ ($j=1,\cdots, K$), its unit normal vector $\uvect{n}_j$ points towards the outside of $\tens{\mathcal{P}}$, and the order of vertices fulfills the right-hand-rule with $\uvect{n}_j$; see Fig. \ref{fig:polyhedra}(b). 
Then, the form factor of $\tens{\mathcal{P}}$ is 
\begin{align}\label{eq:form-factor-polyhedron}
\fn{\tilde{m}}{\vect{k};\tens{\mathcal{P}}} = \frac{-1}{i k^2} \vect{k}\cdot \sum_{j=1}^K \uvect{n}_j\fn{\tilde{m}}{\vect{k};\tens{\Gamma}_j},
\end{align}
where $k\equiv \abs{\vect{k}}$. 
The reader is referred to Ref. \cite{wuttke_form_2017} for derivations of Eqs. \eqref{eq:form-factor_J-gon} and \eqref{eq:form-factor-polyhedron} \footnote{Equations \eqref{eq:form-factor_J-gon} and \eqref{eq:form-factor-polyhedron} are the complex conjugates of the formulas derived in Ref. \cite{wuttke_form_2017} due to the sign convention of Fourier transform.}.

\section{Periodic Network Models}
\label{sec:models}

\begin{figure*}
\includegraphics[width=0.7\textwidth]{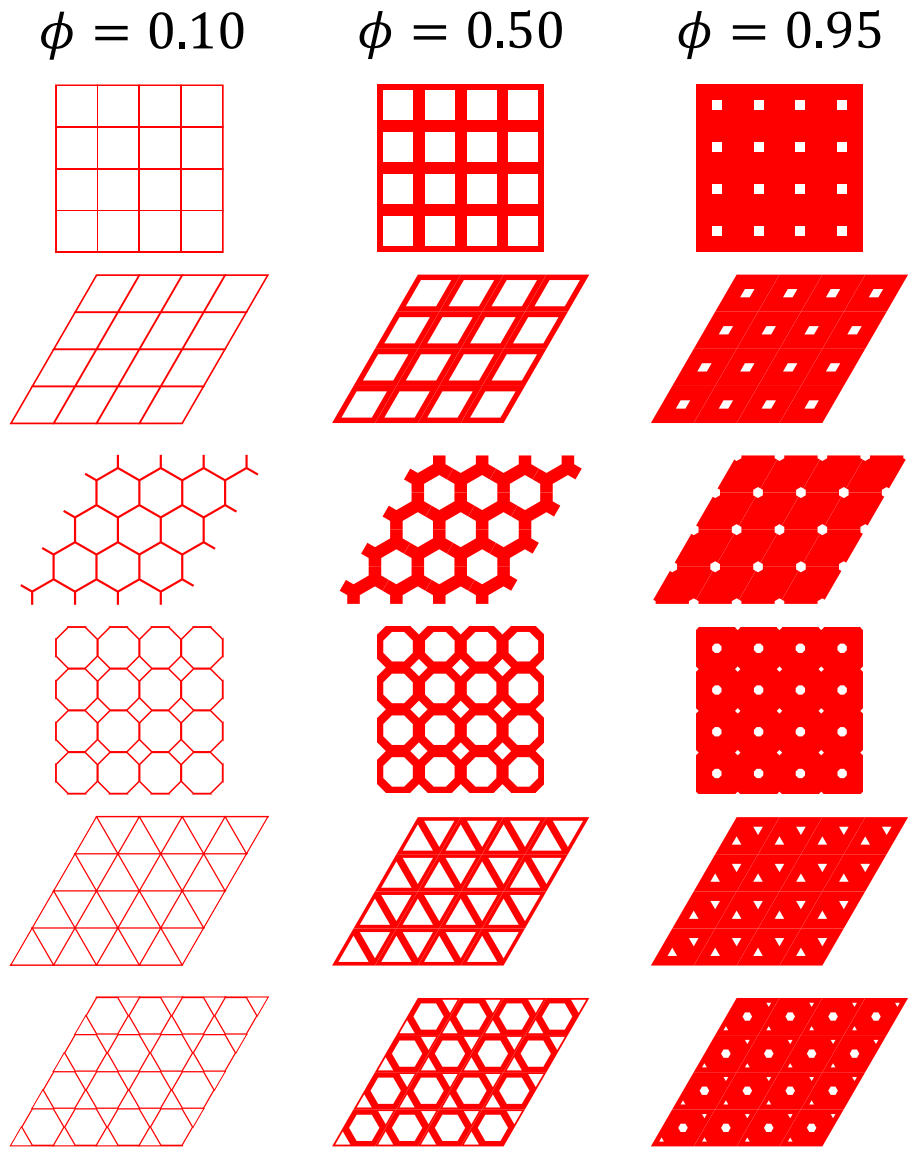}
\caption{
Illustrations of the six different periodic cellular networks considered in this paper: from top to bottom, the square, rhombic, honeycomb, square-octagon, triangular, and kagom\'{e} networks. 
We show each of them at three solid-phase volume fractions: $\phi=0.10$, $\phi=0.50$, and $\phi=0.95$.
Note that these networks can be regarded as periodic point patterns in the limit of $\phi \to 1$. 
\label{fig:illustr-networks}}
\end{figure*}

\begin{figure}
\includegraphics[width=0.45\textwidth]{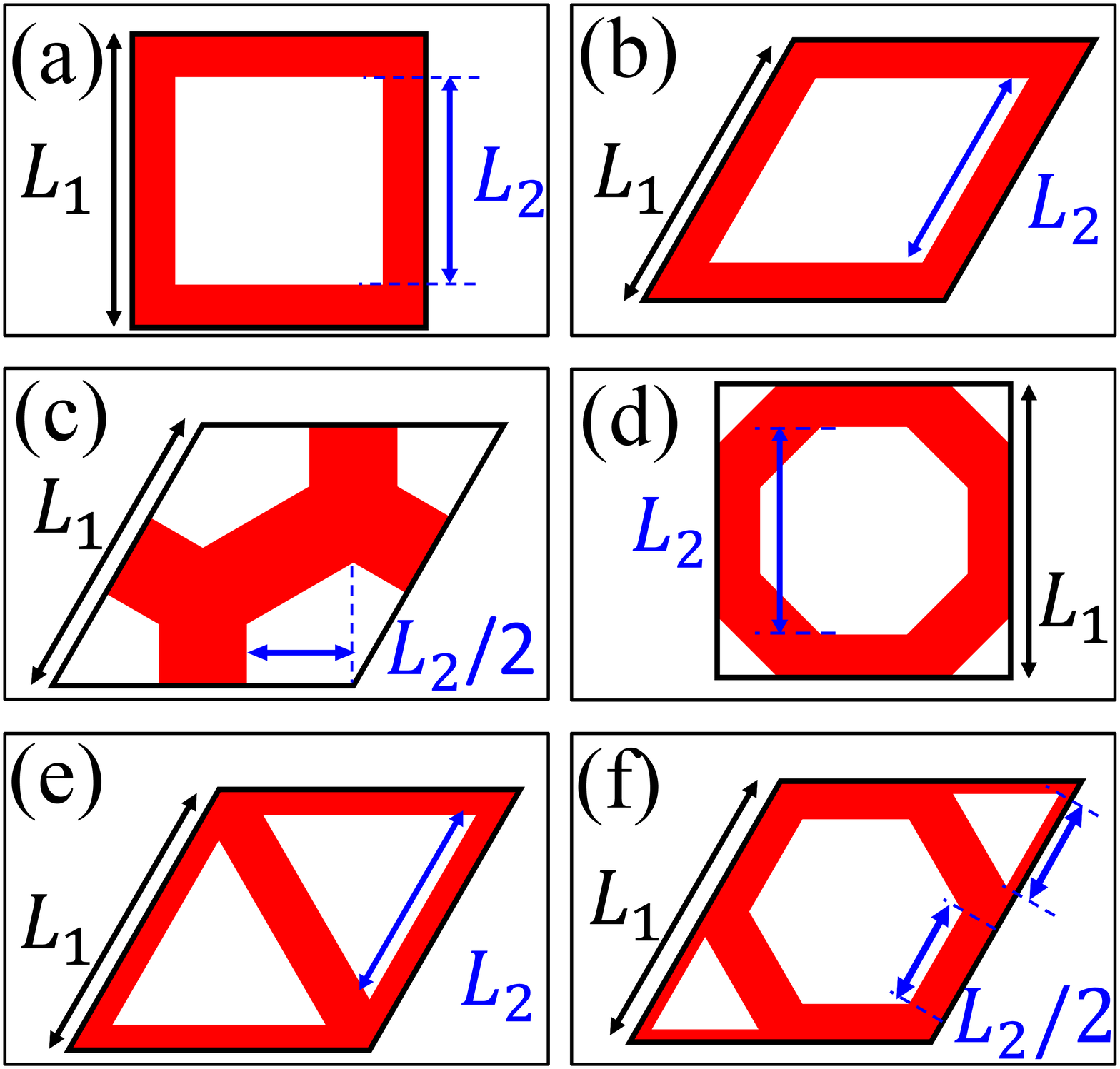}
\caption{Unit cells of two-dimensional periodic networks: (a) square, (b) rhombic, (c) honeycomb, (d) square-octagon, (e) triangular, and (f) kagom\'{e} networks.
While (a) and (d) have square fundamental cells, the rest of networks have rhombic fundamental cells. 
The length parameters $L_1$ and $L_2$ (shown in black and blue arrows, respectively) determine the volume fraction of the solid phase (red regions) $\phi = 1-(L_2/L_1)^2$.
These periodic networks can be treated as packings of polygons defined by the white regions: (a)-(c) can be expressed by a single polygon in the fundamental cells, whereas (d)-(f) needs multiple polygons.   
\label{fig:sch}}
\end{figure}

In this paper, we consider six different periodic networks with
volume fractions that span the entire interval [0,1]: square, rhombic,
honeycomb, square-octagon, triangular, and kagom\'{e} networks.
Figure \ref{fig:illustr-networks} shows each of these networks at small, intermediate, and large solid volume fractions ($\phi=0.10,~0.5, $ and $0.95$, respectively). 
Figure \ref{fig:sch} provides the dimensional parameters for the unit cells, which determine the corresponding solid-phase volume fractions. 
Except for the kagom\'{e} and square-octagon networks, all ``wall" thicknesses are uniform across all volume fractions.
In the cases of the former two structures, the wall thicknesses are uniform for each different polygon but are proportional to their area ratios in order to span all volume fractions in the interval [0,1].
Note that in the limit of $\phi\to 1$, these six network models can be regarded as periodic point patterns.
For example, the square, honeycomb, and triangular networks become the square-lattice, triangular-lattice,and honeycomb crystal, respectively. 


It is noteworthy that these cellular solids can optimize certain effective physical properties.
In the limit $\phi \to 0$, these network structures maximize certain effective transport and elastic properties.
Specifically, all networks maximize the effective conductivity $\sigma_e$ and effective bulk modulus $K_e$ \cite{torquato_effective_1998-1}.
The effective shear modulus $G_e$ is maximized for the triangular network \cite{torquato_effective_1998-1, hyun_effective_2000} as well as the kagom\'e network \cite{christensen_mechanics_2000}.
The triangular and kagom\'e networks are nearly optimal for $\sigma_e$, $K_e$ and $G_e$ over the possible range of volume fractions \cite{hyun_optimal_2002}.
Due to the well-known mechanisms that lead to optimality in the aforementioned networks, we can report here that the rhombic and square-octagon networks maximize the effective conductivity and effective bulk moduli in the limit of $\phi\to0$. 

Many of the periodic networks considered in this paper can be derived from the tessellations associated with certain underlying periodic point configurations.
In order to make contact with the corresponding rank ordering of class I periodic point configurations previously obtained \cite{Torquato2003_hyper, Zachary2009a} and the rank ordering of our two-phase networks, it is instructive here to briefly describe the relationships between the point configurations and their tessellations.
For example, the Voronoi tessellation associated with points arranged on a square lattice yields the square network.
The Voronoi tessellation associated with points arranged on a triangular-lattice yields the honeycomb network.
The Voronoi tessellation associated with points arranged on a honeycomb
crystal yields the triangular network.

%


\section{Periodic and Disordered Packing Models}
\label{sec:packings}

\begin{figure}
\includegraphics[width = 0.45\textwidth]{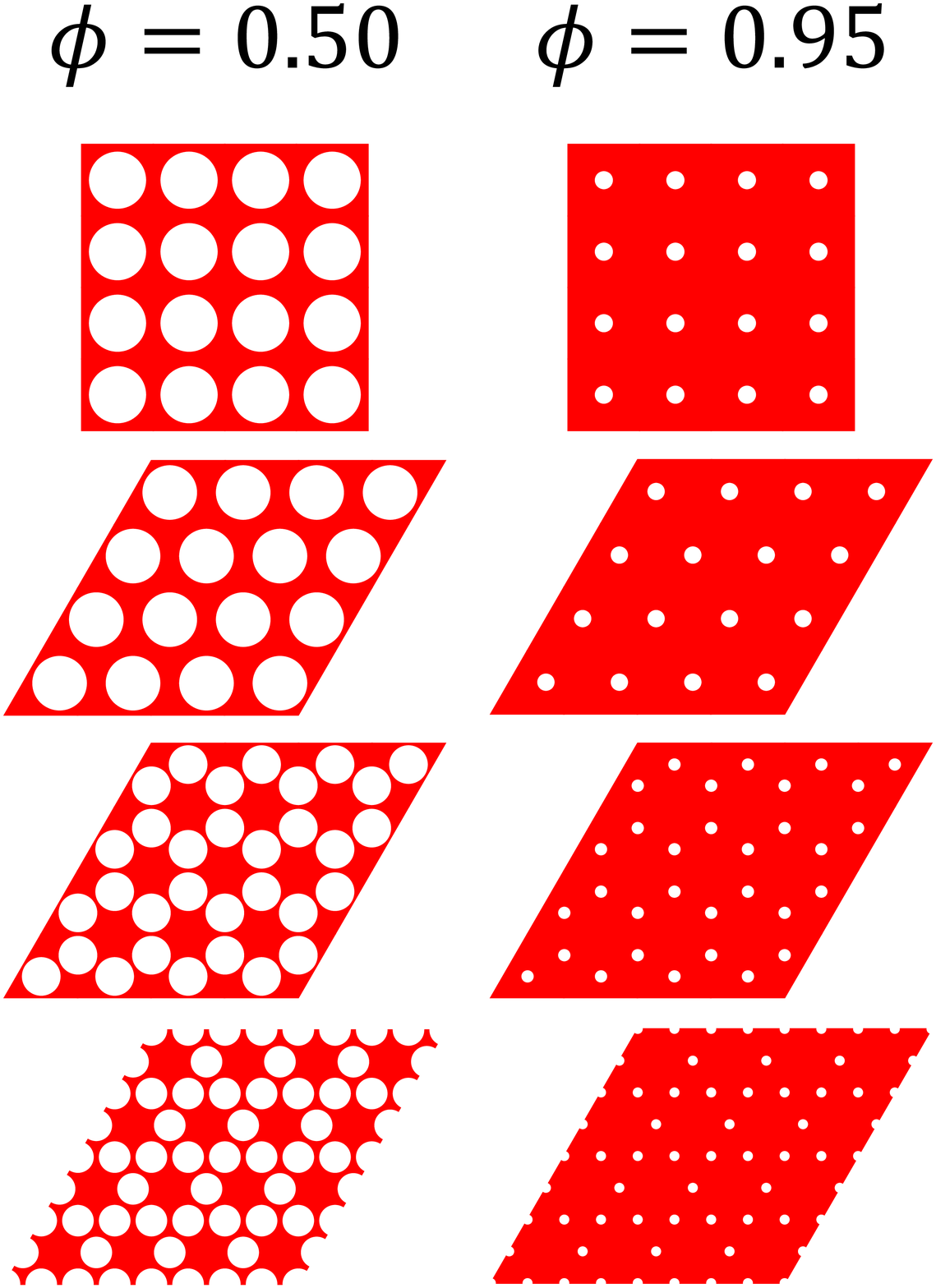}
\caption{Illustrations of the four different models of the two-dimensional periodic dispersions of nonoverlapping identical disks considered in this paper with different solid-phase volume fractions: $\phi=0.50$ and $\phi=0.95$.
From top to bottom, we present dispersions associated with the square and triangular lattices and honeycomb and kagom\'e crystals. 
\label{fig:sch-lattice-packing}
}
\end{figure}

\begin{figure*}
\includegraphics[width = 0.7\textwidth]{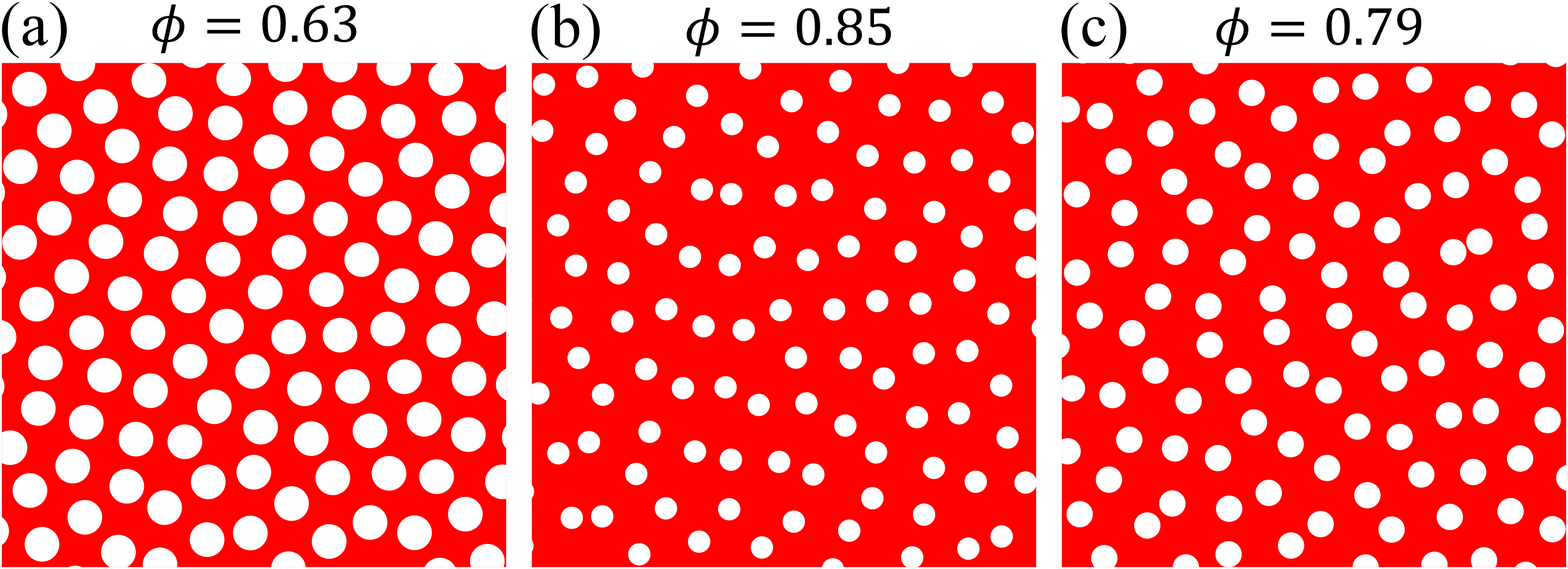}
\caption{
Representative images of the three different models of the two-dimensional disordered/irregular  packings of identical circular disks considered in this paper at different volume fractions: (a) stealthy hyperuniform packing of $\chi=0.49$ and $\phi=0.63$, (b) stealthy hyperuniform packing of $\chi=0.40$ and $\phi=0.85$, and (c) perturbed-lattice packing of $\phi=0.79$.
\label{fig:sch-disordered}
}
\end{figure*}

Here we consider four different two-dimensional dispersions of identical nonoverlapping circular disks on the sites of the triangular and squares lattices as well as the sites of honeycomb and kagom\'{e} crystals (see Fig. \ref{fig:sch-lattice-packing} for illustrations of each of the periodic packings).
We also investigate red different disordered/irregular packings of circular disks: stealthy hyperuniform packings ($\chi=0.49$ and $\chi=0.4$) and perturbed-lattice packings (see Fig. \ref{fig:sch-disordered} for illustrations of each of these packings).

	Stealthy hyperuniform packings of identical particles, which are also class I, are defined by the spectral density vanishing around the origin, i.e., 
	\begin{equation}
	\spD{\vect{k}}=0,~\text{for }0\leq \abs{\vect{k}} \leq K.
	\end{equation}
	Specifically, we first generate stealthy hyperuniform point configurations that include $N$ particles in a periodic fundamental cell $\mathcal{F}$ via the collective-coordinate optimization technique \cite{Uche2004, Batten2008, Zhang2015}.
	We then circumscribe the points by identical nonoverlapping disks \cite{Zhang2016}. 
	For stealthy hyperuniform packings (or point patterns), it is useful to define the $\chi$ parameter, which the ratio of constrained degrees of freedom to total
number of degrees of freedom \cite{Torquato2015_stealthy, Zhang2015}, i.e.,
\begin{equation}
\chi\equiv \frac{\mathcal{M}}{d(N-1)}. \label{eq:chi-value}
\end{equation}
For $0 < \chi < 1/2$, the stealthy hyperuniform point patterns are highly degenerate and disordered, whereas for $1/2< \chi < 1$ they crystallize \cite{Zhang2015}.
	Remarkably, disordered stealthy hyperuniform nonoverlapping spherical obstacles 
(for sufficiently high $\chi$ below 1/2) in a liquid also have nearly maximal effective diffusion coefficients
as well as maximal effective thermal/electrical conductivities for perfectly
insulating inclusions \cite{Zhang2016}.
	
	In this work, we numerically generate 30 different point patterns of $10^4$ particles with $\chi=0.4$ and $\chi=0.49$. Then we determine their corresponding largest fractions of space covered by the disks, which is equivalent to smallest possible solid-phase volume fraction $\phi_{\min}$, equal to about $0.153$ (i.e., $\phi \geq \phi_{\min}=0.85$) and $0.377$ (i.e., $\phi \geq \phi_{\min}=0.63$), respectively.

	We also generate perturbed-lattice packings by independently displacing each point of a square lattice by a random vector that is uniformly distributed in a closed square $[-a/2, a/2]^2$ \footnote{
	Such point configurations are characterized by a structure factor that contains a diffuse (disordered) contribution and Bragg-diffraction (long-range order) contribution \cite{Gabrielli2004, Jaeuk2018} and therefore cannot be considered to be truly disordered, as measured quantitatively by the $\tau$ order metric \cite{klatt_cloaking_2020}.
	For this reason, we reserve the term {\it irregular} for such perturbed-lattice packings.}. We then circumscribe the resulting points by identical nonoverlapping disks. 
	The resulting point pattern (or packing) is class I hyperuniform; see Refs. \cite{Gabrielli2004, Jaeuk2018, klatt_cloaking_2020} for details. 
	In this work, we numerically generate 50 configurations of $10^4$ particles and $a=0.48$. We find that their largest possible fraction of space covered by the disks, which is equivalent to the smallest possible solid-phase volume fraction $\phi_{\min}$, is around $0.213$ (i.e., $\phi \geq \phi_{\min}=0.79$).

\section{Characteristic Length Scales}
\label{sec:length-scale}

When ranking class I hyperuniform systems according to the hyperuniformity order metric $\overline{B}_V$ ($\overline{B}_N$for the point-configuration counterparts \cite{Torquato2003_hyper, Zachary2009a}), it is critical to choose an appropriate characteristic length scale $D$ because these order metrics depend on $D$, as we noted in Sec. \ref{sec:asy}.
In the case of hyperuniform point patterns in $\R^d$, it is natural to choose $D=\rho^{-1/d}$, where $\rho$ is the number density of points.
However, the choice of a length scale in the case of two-phase media is highly nontrivial because the geometries and topologies of the phases are generally much richer and more complex than point configurations.
Indeed, there are an infinite number of ways of decorating a point configuration to produce two-phase media, all of which cannot be universally characterized.

In this paper, we consider and evaluate several possible choices for the length scale $D$ according to the following three criteria:
(i) $D$ must be defined for general two-phase media, (ii) $D$ must be independent of the choice of phase, and (iii) the associated order metric $\overline{B}_V$ must be a finite number for any volume fraction.
Seemingly obvious choices for $D$, including the size of a fundamental cell for periodic systems or the mean nearest-distance for disordered or irregular packings, fail to meet the criteria (i) and (ii).
There are several candidates that satisfy the criterion (i), such as the mean chord length of one phase (i.e., the expected length of line segments in the phase between the intersections of an infinitely long line with the two-phase interface \cite{underwood_quantitative_1970, torquato_chord-length_1993, Torquato_RHM}). 
However, criteria (ii) and (iii) immediately eliminate the mean chord length of an individual phase.
The resulting $\overline{B}_V$ diverges at either $\phi=0$ or $\phi=1$.
Averages based on the mean chord length for each phase, such as the arithmetic and geometric means, satisfy all criteria.
One such example is the inverse of the specific surface (i.e., the mean interface area per volume) $1/s$, which turns out to be directly proportional to the arithmetic mean of the mean chord length $\ell_C^{(i)}$ of both phases, i.e., $1/s = (\ell_C^{(1)}+\ell_C^{(2)})/\pi$ \cite{Torquato_RHM}.
Explicit formulas for the specific surface $s$ of all models considered in the paper are provided in in Appendix \ref{sec:s}. 
Henceforth, we employ $D=s^{-1}$.

\section{Results}
\label{sec:results}

\begin{figure*}
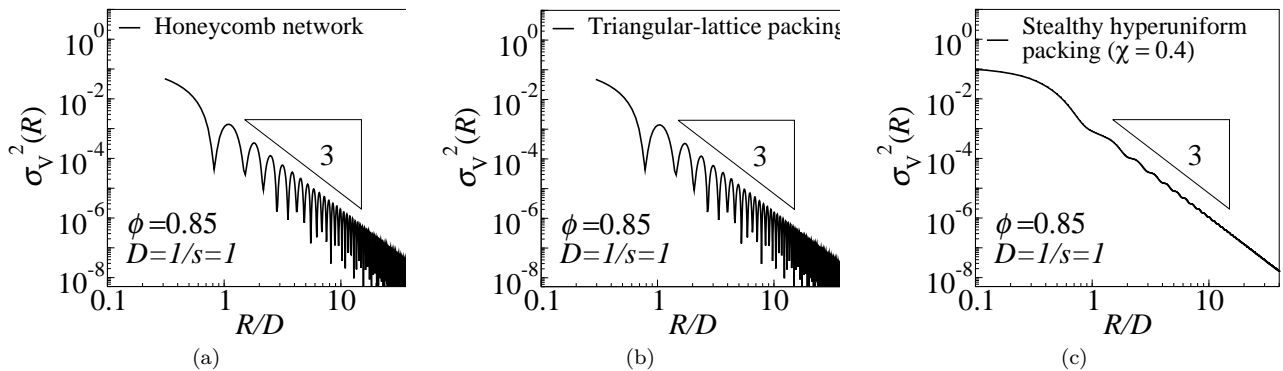

\subfloat[]{
\includegraphics[width=0.3\textwidth]{fig6a.eps}}
\hspace{5pt}
\subfloat[]{
\includegraphics[width=0.3\textwidth]{fig6b.eps}}
\hspace{5pt}
\subfloat[]{
\includegraphics[width=0.3\textwidth]{fig6c.eps}}
\caption{Log-log plots of the local volume-fraction variances $\vv{R}$ of two-dimensional ordered and disordered cellular solids at a selected solid-phase volume fraction $\phi=0.85$: (a) honeycomb network, (b) triangular-lattice disk packing, and (c) stealthy hyperuniform packings of $\chi = 0.4$.
The first two models are periodic structures, whereas the last is a disordered one.
Here we take the inverse of the specific surface $1/s$ to be unity, i.e., $D=1/s=1$. 
\label{fig:vv}}
\end{figure*}

We consider two-dimensional ordered and disordered two-phase media, shown in Figs. \ref{fig:sch}, \ref{fig:sch-lattice-packing}, and \ref{fig:sch-disordered} with taking the inverse of the specific surface as the characteristic length scale, i.e., $D=1/s=1$. 
Figure \ref{fig:vv} shows log-log plots of local volume-fraction variances $\vv{R}$ as a function of window radius $R$ at a solid-phase volume fraction$\phi=0.85$ for honeycomb network, triangular-lattice packing, and disordered stealthy hyperuniform packings (with $\chi = 0.40$).
The variances for the periodic models and the disordered example are obtained from Eq. \eqref{eq:vv-perio-network} and the MC method, respectively.
For all models considered here, $\vv{R}$ globally decays as fast as $R^{-3}$ in the large-$R$ regime, which are of class I hyperuniformity [cf. Eq. \eqref{eq:classes}], and fluctuates on ``microscopic" length scales, which in the case of periodic structures, are associated with the spacing of the underlying Bravais lattice. 

\begin{figure*}
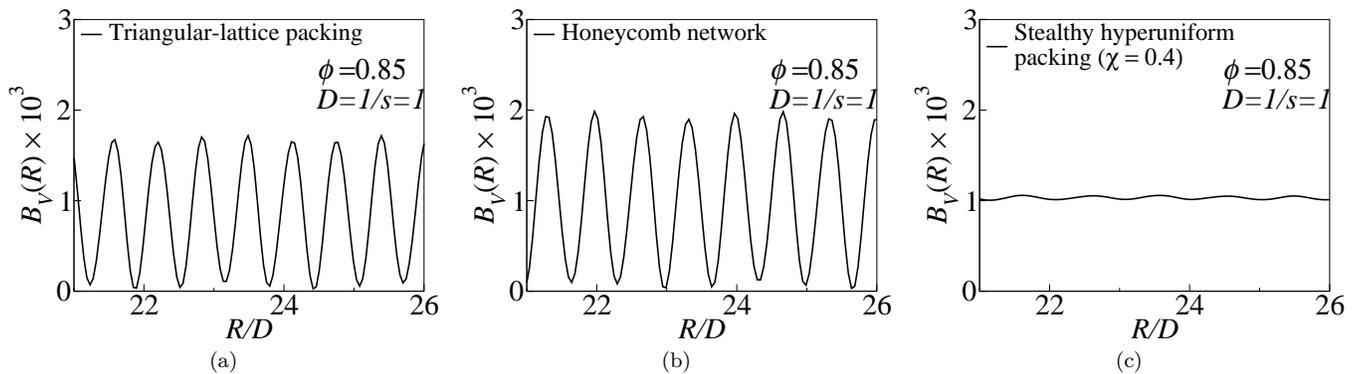

\subfloat[]{\includegraphics[width=0.32\textwidth]{fig7a.eps}}
\hspace{5pt}
\subfloat[]{\includegraphics[width=0.32\textwidth]{fig7b.eps}}
\hspace{5pt}
\subfloat[]{\includegraphics[width=0.32\textwidth]{fig7c.eps}}
\caption{Surface-area coefficient $\fn{B_V}{R}$ as a function of window radius $R$ of two-dimensional ordered and disordered cellular solids at a selected solid-phase volume fraction $\phi=0.85$, as per Fig. \ref{fig:vv}. 
\label{fig:Bv}}
\end{figure*}

We plot the surface-area coefficients $\fn{B_V}{R}$ for the models considered in Fig. \ref{fig:vv} to more closely investigate such local fluctuations of $\vv{R}$; see Fig. \ref{fig:Bv}.
As pointed earlier, $\fn{B_V}{R}$ oscillates around an average value $\overline{B}_V$. 
For disordered systems [shown in Fig. \ref{fig:Bv}(c)], such oscillations typically decay as $R$ increases, whereas for periodic networks, the amplitude of the oscillations does not decrease, even in the limit of $R\to\infty$.



\begin{figure*}
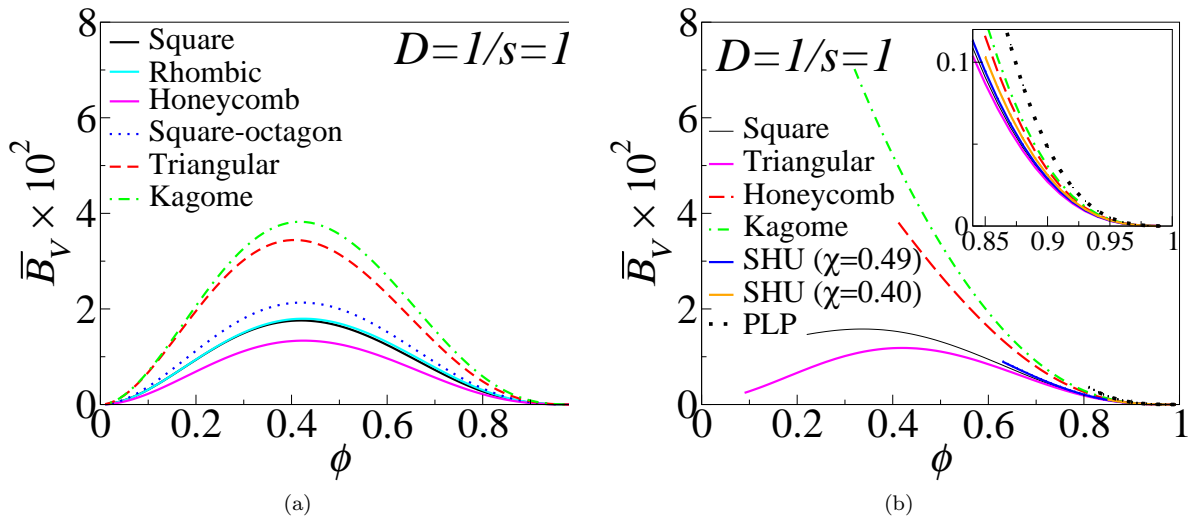

\subfloat[]{\includegraphics[width=0.43\textwidth]{fig8a.eps}}
\hspace{5pt}
\subfloat[]{\includegraphics[width=0.43\textwidth]{fig8b.eps}}
\caption{  
Asymptotic values of the surface-area coefficient $\overline{B}_V$ as a function of the solid-phase volume fraction $\phi$ for (a) two-dimensional periodic networks and (b) ordered and disordered disk packings.
We take the length scale as $D=1/s=1$, where $s$ is the specific surface. 
In (b), SHU and PLP stand for the stealthy hyperuniform packing and perturbed-lattice packing, respectively.
The inset in (b) is a magnification of the larger panel. 
\label{fig:coefficients}}
\end{figure*}

Figure \ref{fig:coefficients} shows how the hyperuniformity order metric $\overline{B}_V$ for two-dimensional two-phase media varies with the volume fraction $\phi$.
For periodic networks or disk packings, $\overline{B}_V$ is evaluated from Eq. \eqref{eq:Bv-asy-period}, whereas the disordered/irregular counterparts are evaluated by applying the running average associated with \eqref{eq:Bv-asy} to the MC results. 
For periodic networks, where $\phi$ can span from 0 to 1, as shown in Fig. \ref{fig:coefficients}(a), $\overline{B}_V$ exhibits the following three common characteristics: (i) it vanishes trivially at $\phi=0$ and $\phi=1$, (ii) it is proportional to $\phi^2$ for small $\phi$, and (iii) it has a maximum at around $\phi=0.4$.
By contrast, for periodic or disordered disk packings, $\overline{B}_V$ trivially vanishes at $\phi=1$, but the other characteristics are not observed; see Fig. \ref{fig:coefficients}(b).
We first investigate the rankings of $\overline{B}_V$ for periodic networks shown in Fig. \ref{fig:coefficients}(a) and those for disk packings in Fig. \ref{fig:coefficients}(b) separately and then discuss the rankings for all models. 
Among the considered periodic networks, honeycomb and kagom\'e ones achieve the minimum and maximum values of $\overline{B}_V$, respectively, at a given value of volume fraction $\phi$.
For the six network models, the values $\overline{B}_V$ increases from  honeycomb, square, rhombic, square-octagon, triangular, to kagom\'e ones. 
Note that the ranking for the honeycomb, square, and triangular networks are consistent with the ranking of the corresponding metrics for the point counterparts of these three network models (triangular lattice, square lattice, and honeycomb crystals, respectively, as discussed in Sec. \ref{sec:models}) given in Ref. \cite{Torquato2003_hyper, Zachary2009a}.
Moreover, we note that periodic networks with a single void region in the fundamental cell (honeycomb, square, and rhombic) tend to be more ordered (i.e., smaller $\overline{B}_V$) than those with multiple void regions in the fundamental cell (square-octagon, triangular, and kagom\'e).

Figure \ref{fig:coefficients}(b) shows that among the periodic disk packings, the triangular and kagom\'e packings achieve the minimum and maximum values of $\overline{B}_V$, respectively. 
Considering all models of disk packings, the values of $\overline{B}_V$ at a given volume fraction $\phi$ increases from triangular, square, disordered stealthy ($\chi=0.49$), disordered stealthy ($\chi=0.4$), honeycomb, kagom\'e, to perturbed-lattice.
Similar to the case of periodic networks, the Bravais-lattice packings (triangular and square) are more ordered than non-Bravais-lattice packings (honeycomb and kagom\'e).
Importantly, such a ranking of disk packings is identical to the rankings of the point counterparts that were reported in Ref. \cite{Zachary2009a, Torquato2018_review}, i.e., triangular, square, disordered stealthy ($\chi=0.496$), disordered stealthy ($\chi=0.402$), honeycomb, and kagom\'e.

In the discussion above, we consider the rankings for periodic network models [Fig. \ref{fig:coefficients}(a)] and those for disk packings [Fig. \ref{fig:coefficients}(b)] separately.
There, the rankings for the models in each class does not change as the solid-phase volume fraction $\phi$ is varied. 
However, when rankings all models in both periodic networks and periodic and disordered disk packings, the resulting rankings can change with $\phi$, and hence the volume fraction $\phi$ should be specified. 
For this purpose, we tabulate $\overline{B}_V$ for the periodic networks, periodic disk packings, and disordered disk packings at selected values of the solid-phase volume fraction $\phi$ in Tables \ref{tab:order-1}, \ref{tab:order-2}, and \ref{tab:disordered}, respectively.
From Tables \ref{tab:order-1} and \ref{tab:order-2}, we immediately see that while the triangular disk packing is generally more ordered than the honeycomb network, their rankings change for $\phi \lesssim 0.1$.
As shown in Tables\ref{tab:order-1}-\ref{tab:disordered}, among all considered models at $\phi=0.85$, the triangular-lattice packing and perturbed-lattice packing have the smallest and highest values of $\overline{B}_V$, respectively.
We also note that at $\phi=0.85$, the triangular- and square-lattice packingshave lower order metrics than their network counterparts (i.e., honeycomb and square networks, respectively).
This implies that the length scale $D=1/s=1$ penalizes the order metric $\overline{B}_V$ of a packing of nonspherical particles compared to the corresponding sphere packing.

\begingroup
\begin{table*}
\caption{
Hyperuniformity order metric $\overline{B}_V$ of the six models of two-dimensional periodic cellular networks at various values of volume fraction $\phi$; see Fig. \ref{fig:sch}.
The quantities are computed by Eq. \eqref{eq:Bv-asy-period} and taking the characteristic length scale to be the inverse of the specific surface, i.e., $D=1/s=1$.
Note that for a given value of $\phi$, $\overline{B}_V$ increases from top to bottom. 
\label{tab:order-1}}
\begin{tabular}{c|c|c|c|c|c|c}
\hline
$\phi$&		0.1&	0.25&	0.4&	0.55&	0.7&	0.85	\\
\hline
Honeycomb&	2.2570$\times10^{-3}$&	9.2268$\times10^{-3}$&	1.3279$\times10^{-2}$&	1.1351$\times10^{-2}$&	5.6546$\times10^{-3}$&	1.0020$\times10^{-3}$\\
Square&		3.1288$\times10^{-3}$&	1.2433$\times10^{-2}$&	1.7512$\times10^{-2}$&	1.4717$\times10^{-2}$&	7.2338$\times10^{-3}$&	1.2693$\times10^{-3}$\\
Rhombic&	3.1320$\times10^{-3}$&	1.2519$\times10^{-2}$&	1.7846$\times10^{-2}$&	1.5301$\times10^{-2}$&	7.7541$\times10^{-3}$&	1.4222$\times10^{-3}$\\
Square-octagon&	3.7286$\times10^{-3}$&	1.4953$\times10^{-2}$&	2.1225$\times10^{-2}$&	1.7943$\times10^{-2}$&	8.8514$\times10^{-3}$&	1.5538$\times10^{-3}$\\
Triangular&	6.7514$\times10^{-3}$&	2.5523$\times10^{-2}$&	3.4413$\times10^{-2}$&	2.7739$\times10^{-2}$&	1.3071$\times10^{-2}$&	2.1926$\times10^{-3}$\\
Kagom\'e&	6.9411$\times10^{-3}$&	2.7262$\times10^{-2}$&	3.8114$\times10^{-2}$&	3.1837$\times10^{-2}$&	1.5553$\times10^{-2}$&	2.7074$\times10^{-3}$\\
\hline
\end{tabular}
\end{table*}
\endgroup

\begingroup
\begin{table*}
\caption{
Hyperuniformity order metric $\overline{B}_V$ of the four models of two-dimensional periodic disk packings at various values of volume fraction $\phi$; see Fig. \ref{fig:sch-lattice-packing}.
The quantities are computed by Eq. \eqref{eq:Bv-asy-period} and taking the characteristic length scale to be the inverse of the specific surface, i.e., $D=1/s=1$.
Note that for a given value of $\phi$, $\overline{B}_V$ increases from top to bottom. 
\label{tab:order-2}}
\begin{tabular}{c|c|c|c|c|c|c}
\hline
$\phi$	&	0.1&			0.25&			0.4&			0.55&			0.7&			0.85\\
\hline
Triangular&	2.7700$\times 10^{-3}$&8.5512$\times 10^{-3}$&	1.1794$\times 10^{-2}$&	9.9491$\times 10^{-3}$&	4.9261$\times 10^{-3}$&	8.6942$\times 10^{-4}$\\
Square&		$-$		&	1.5130$\times 10^{-2}$&	1.5402$\times 10^{-2}$&	1.1455$\times 10^{-2}$&	5.3311$\times 10^{-3}$&	9.0849$\times 10^{-4}$\\
Honeycomb&	$-$		&	$-$		&	$-$		&	2.1331$\times 10^{-2}$&	7.9380$\times 10^{-3}$&	1.1566$\times 10^{-3}$\\
Kagom\'e&		$-$		&	$-$		&	5.2200$\times 10^{-2}$&	2.5962$\times 10^{-2}$&	9.0773$\times 10^{-3}$&	1.2592$\times 10^{-3}$\\
\hline
\end{tabular}
\end{table*}
\endgroup

\begingroup
\begin{table*}
\caption{
Hyperuniformity order metric $\overline{B}_V$ of the three models of two-dimensional disordered disk packings at a selected volume fraction $\phi=0.85$ and their respective lowest volume fractions $\phi_{\min}$; see Fig. \ref{fig:sch-disordered}.
The values of $\phi_{\min}$ for 
{various models are provided in Fig. \ref{fig:sch-disordered}.}
The quantities are computed by the MC procedure and Eq. \eqref{eq:Bv-asy} and taking the characteristic length scale to be the inverse of the specific surface, i.e., $D=1/s=1$.
The uncertainties are estimated from the statistical errors in the estimation of $\vv{R}$.
\label{tab:disordered}}
\begin{tabular}{c|c|c}
\hline
Model	& $\overline{B}_V$ at $\phi_{\min}$	& $\overline{B}_V$ at $\phi=0.85$\\
\hline
Stealthy hyperuniform packings ($\chi=0.49$) & 8.9655(6)$\times 10^{-3}$& 9.4349(6)$\times 10^{-4}$	\\
Stealthy hyperuniform packings ($\chi=0.4$) & 1.0313(1)$\times 10^{-3}$ & 1.0313(1)$\times 10^{-3}$	\\
Perturbed-lattice packing & 3.6034(3)$\times 10^{-3}$&	1.7298(1)$\times 10^{-3}$\\
\hline
\end{tabular}
\end{table*}
\endgroup

\section{Conclusions and Discussion}
\label{sec:conclusion}

In this work, we took initial steps to characterize a restricted subset of class I hyperuniform two-phase media in two dimensions by ascertaining their local volume-fraction variances $\vv{R}$ and the associated hyperuniformity order metrics $\overline{B}_V$.
These models include a variety of different periodic cellular networks, periodic packings, and disordered/irregular packings, some of which maximize their effective transport and elastic properties \cite{torquato_effective_1998-1,
hyun_effective_2000,
christensen_mechanics_2000,
hyun_optimal_2002,
Zhang2016}.
Using the estimated $\overline{B}_V$ and a judicious choice for a length scale to make it dimensionless (as discussed below), we ranked these class I models of two-phase media according to the degree to which they suppress large-scale volume-fraction fluctuations. 
Among the periodic networks, the honeycomb and kagom\'e networks always achieve the lowest and highest $\overline{B}_V$, respectively, and the rankings do not change as the solid-phase volume fraction $\phi$ varies. 
Similarly, the rankings for disk packings also do not change with $\phi$. The triangular-lattice packings (whose Voronoi tessellations are honeycomb networks) and the perturbed-lattice packings have the minimum and maximum values of $\overline{B}_V$, respectively.
Not surprisingly, however, the overall rankings for both network and packing models with their distinctly different geometries and topologies are difficult to unscramble because they change with $\phi$.
Nonetheless, we summarize these rankings by making two general observations. 
First, the rankings for packings of identical disks are consistent with those of the point-configuration order metric $\overline{B}_N$ corresponding to their underlying point patterns \cite{Torquato2003_hyper, Zachary2009a, Torquato2018_review} at any considered volume fraction $\phi$. 
Second, for both periodic networks and periodic packings with the same underlying Bravais lattice, the structures with smaller specific surfaces have lower values of $\overline{B}_V$.
We note that the second observation is generally true, with a few notable exceptions in which the volume fraction of the solid phase becomes so low that the disks are nearly in contact with one another. 
Specifically, among all models considered in this work, triangular-lattice packing has the minimal $\overline{B}_V$ for all solid-phase volume fraction greater 0.1.
Otherwise, the honeycomb networks record the smallest value

When establishing these rankings according to $\overline{B}_V$, it is crucial to determine a characteristic length scale $D$ to make $\overline{B}_V$ dimensionless, which is highly nontrivial due to the need to account for a wide spectrum of two-phase structures.
Among various possibilities, we chose the inverse of the specific surface $1/s$ as the length scale $D$ by considering the three criteria of generality, phase-independence, and boundedness of the associated $\overline{B}_V$.
This choice is also reasonable in that $s$ is easy to compute.
Furthermore, it is one of the Minkowski functionals (i.e., volume, surface area, integrated mean curvature, and Euler number), which are fundamental shape descriptors that have been widely used in various applications \cite{schroder-turk_minkowski_2011, klatt_characterization_2014}.
The integrated mean curvature might also serve as a choice of the length scale $D$.
Although we have made a specific choice $D=s^{-1}$, we note that one can easily convert our results for $\overline{B}_V$ tothe corresponding quantity for any another length scale $D=\ell$ by use of the relation 
\begin{equation}
\overline{B}_V \vert_{D=\ell} = \overline{B}_V /(s \ell)^{d+1},
\end{equation}
where $d$ is the space dimension.

Our study lays the theoretical foundation to establish
the hyperuniformity order metrics of more general two-phase systems.
Towards this end, one needs to develop methods to estimate $\vv{R}$ for a wider class of hyperuniform two-phase media than what can be handled by the methods used in this work, such as labyrinth-like patterns associated with spinodal decomposition \cite{Ma2017}. 
Such a development will also be beneficial in detecting (effective) hyperuniformity of relatively small systems, in which the asymptotic analysis of $\vv{R}$ [cf. \eqref{eq:HU-condition2}] is more reliable than the spectral-density condition \eqref{eq:HU_condition} \cite{dreyfus_diagnosing_2015}.
Further studies in three and higher dimensions will be helpful in determining whether $\overline{B}_V$ scaled by $D=1/s$ is a robust order metric.
It would also be interesting to know whether the two-phase counterpart of the decorrelation principle \cite{torquato_random_2006, zachary_high-dimensional_2011} for disordered two-phase media could be observed as the space dimension increases.

It will be of interest to determine whether hyperuniformity of fluctuations associated with the two-phase interface \cite{Torquato2016_gen} leads to the same rank ordering as for volume-fraction fluctuations for the models considered in this investigation.
Another promising avenue for future study is the construction of two-phase structures with a prescribed value of $\overline{B}_V$.
This problem can be regarded as a type of Fourier-space based inverse design procedure \cite{Chen2017}, in which Eq. \eqref{eq:Bv-asy-period} is taken as the objective function.
Such a procedure can be employed in discovering new types of periodic structures with a specified value of $\overline{B}_V$.
The algorithm developed to solve this problem would also provide a tool for determining whether the triangular-lattice disk packing, which we demonstrated minimizes $\overline{B}_V$ among the considered models, is a global minimizer for $\overline{B}_V$ among a larger class of models.
An interesting question is what physical properties are optimized by the global minimizer of $\overline{B}_V$ under certain constraints.



%
%
%
%

%
%
%
%

%
%
%
%
%
%
%
%
%
%
%
%

{\appendix

\begin{table}[h!]
\caption{Formulas for the specific surface $s$ for all models considered in this work.
For periodic networks and disk packings, the specific surface can be expressed as $s = C\sqrt{1-\phi}/L_1$, where $L_1$ is a length parameter of the unit cells; see Fig. \ref{fig:sch-disordered}.
For any disordered/irregular disk packing of the number density $\rho$, the specific surface is written as $s = C\sqrt{1-\phi} \rho^{1/2}$. 
\label{tab:conversion-factor}
}
\begin{tabular}{c | c | c}
\hline
\multicolumn{2}{c|}{Models}& $C$\\
\hline
\multirow{6}{*}{Periodic networks}
	&Honeycomb	&	$4$\\
	&Square		&	$4$\\
	&Rhombic	&	$8/\sqrt{3}$\\
	&Square-octagon&	$12/(1+\sqrt{2})$\\	&Triangular	&	$4\sqrt{3}$\\
	&Kagom\'e	&	$4\sqrt{3}$\\
	\hline
\multirow{4}{*}{Periodic disk packings}
	&Triangular	&	$2\sqrt{2\pi}/3^{1/4}$\\
	&Square		&	$2\sqrt{\pi}$\\
	&Honeycomb	&	$4\sqrt{\pi}/3^{1/4}$\\
	&Kagom\'e	&	$2\sqrt{2\pi}/3^{1/4}$\\
\hline
\multicolumn{2}{c|}{Disordered disk packing } &	$2\sqrt{\pi}$ \\
\hline		
\end{tabular}
\end{table}

\section{Specific Surface for Various Models of Two-Phase Media}\label{sec:s}

Table \ref{tab:conversion-factor} provides the formulas for the specific surface $s$ for all two-dimensional models of class I hyperuniform two-phase media considered in this work.
We note that all disk packings, ordered or not, have the same specific surface if they are at the same number density $\rho$.

}

\begin{acknowledgments}

We thank M. Klatt, C. Maher, and T. M. Middlemas for very helpful discussions.
The authors gratefully acknowledge the
support of the Air Force Office of Scientific Research Program on
Mechanics of Multifunctional Materials and Microsystems under
award No. FA9550-18-1-0514.

\end{acknowledgments}

%

\end{document}